\documentclass{aa}  

\usepackage{graphicx}
\usepackage{mathabx}
\usepackage{txfonts}
   \newcommand{\unit}[1]{\ensuremath{\, \mathrm{#1}}}
   \newcommand{\zeustw}{{\sc zeustw}}          % code names
   \newcommand{\krome}{{\sc krome}}          % code names
   \newcommand{\Fig}[1]{Fig.~\ref{fig:#1}}    % Fig. reference
   \newcommand{\Figure}[1]{Figure~\ref{fig:#1}}    % Figure reference

\begin{document}

   \title{Ionization: a possible explanation for the difference of mean disk sizes in star-forming regions}

   \author{M. Kuffmeier
          \inst{1}\fnmsep\thanks{International Postdoctoral Fellow of Independent Research Fund Denmark (IRFD)}
          \and
          B. Zhao\inst{2}
          \and
          P. Caselli \inst{2}
          }

   \institute{Zentrum für Astronomie der Universität Heidelberg, Institut für Theoretische Astrophysik, Albert-Ueberle-Str. 2, 69120 Heidelberg
              \email{kuffmeier@uni-heidelberg.de}
             \and
  Max-Planck Institute for Extraterrestrial Physics, Giessenbachstrasse 1, 85748 Garching 
%             \and
%  Department of Astronomy, University of Virginia, Charlottesville, VA 22904, USA
             }

   \date{Received \today}

% \abstract{}{}{}{}{} 
% 5 {} token are mandatory
 
  \abstract
   {Surveys of protoplanetary disks in star-forming regions of similar age revealed significant variations in average disk mass in some regions. For instance, disks in the Orion Nebular Cluster (ONC) and Corona Australis (CrA) are on average smaller than disks observed in Lupus, Taurus, Chamaeleon I, or Ophiuchus.}
   {In contrast to previous models that studied the truncation of disks at a late stage of their evolution, we investigate whether disks may already be born with systematically smaller disk sizes in more massive star-forming regions as a consequence of higher ionization rates. }
   {Assuming various cosmic-ray ionization rates, we computed the resistivities for ambipolar diffusion and Ohmic dissipation with a chemical network, and performed 2D nonideal magnetohydrodynamical protostellar collapse simulations. }
   {A higher ionization rate leads to stronger magnetic braking, and hence to the formation of smaller disks. 
   Accounting for recent findings that protostars act as forges of cosmic rays and considering only mild attenuation during the collapse phase, we show that a high average cosmic-ray ionization rate in star-forming regions such as the ONC or CrA can explain the detection of smaller disks in these regions. }
   {Our results show that on average, a higher ionization rate leads to the formation of smaller disks. 
   Smaller disks in regions of similar age can therefore be the consequence of different levels of ionization, and may not exclusively be caused by disk truncation through external photoevaporation. 
   We strongly encourage observations that allow measuring the cosmic-ray ionization degrees in different star-forming regions to test our hypothesis.    }

   \keywords{magnetohydrodynamics (MHD), protoplanetary disks, stars: low-mass, magnetic fields, surveys, (ISM:) cosmic rays
                }

   \maketitle
   
\section{Introduction}

In the past years, surveys of young stellar objects (YSOs) of Class II in several star-forming regions have helped significantly to better constrain the size and mass distribution of disks based on dust continuum observations.
A comparison of the obtained average disk sizes and disk masses of the surveys shows significant differences. 
There seems to be a trend that disks in older star-forming regions are typically smaller than disks in younger star-forming regions. 
The oldest star-forming region for which dust masses of disks were obtained is Upper Sco \citep{Barenfeld2016} with an age of 5 -- 10 Myr. 
The average dust mass is $M_{\rm dust} = 5\pm3$ M$_{\Earth}$.
This is significantly lower than the dust masses of disks derived for younger star-forming regions of age $\sim$1 -- 3 Myr, such as Lupus \citep{Ansdell2016}, Chamaeleon I \citep{Pascucci2016}, and Taurus \citep{Long2018}, where the average dust content is derived to be more than 10 M$_{\Earth}$. 
All of the above dust masses were computed using the same method and assumptions \citep[see table 4 in][]{Cazzoletti2019}, which allows a fair comparison of the obtained masses.
The age of the $\sigma$-Orionis region is $\sim$3 -- 5 Myr \citep{Oliveira2002,Oliveira2004}, and hence older than the young star-forming regions Lupus, Chamaeleon I, Ophiuchus, and Taurus, but younger than Upper Sco.
\citet{Ansdell2017} determined the average mass in $\sigma$-Orionis to be $7\pm1$ M$_{\Earth}$, which is lower than for the young regions, but higher than for the older region Upper Sco. 
In agreement with this trend, the dust mass of Class II disks is about five times lower than the dust mass of Class I objects in Ophiuchus, as shown by \citet{Williams2019}, who compared the results of the sample of fainter (and generally more evolved star-disk systems) with the previous sample of brighter disks \citep{Cieza2019}.
Similarly, the dust masses in the $\sim 1$ Myr old Orion Molecular Cloud 2 (OMC-2) are similar to the dust masses measured for Lupus and Taurus, which have similar ages \citep{vanTerwisga2019}.
This correlation of decreasing dust mass with increasing age is consistent with the theory of a radial drift of dust grains in gaseous disks \citep{Whipple1972, Weidenschilling1977,Birnstiel2010}. 

\begin{figure}
        \centering
    \includegraphics[width=0.49\textwidth]{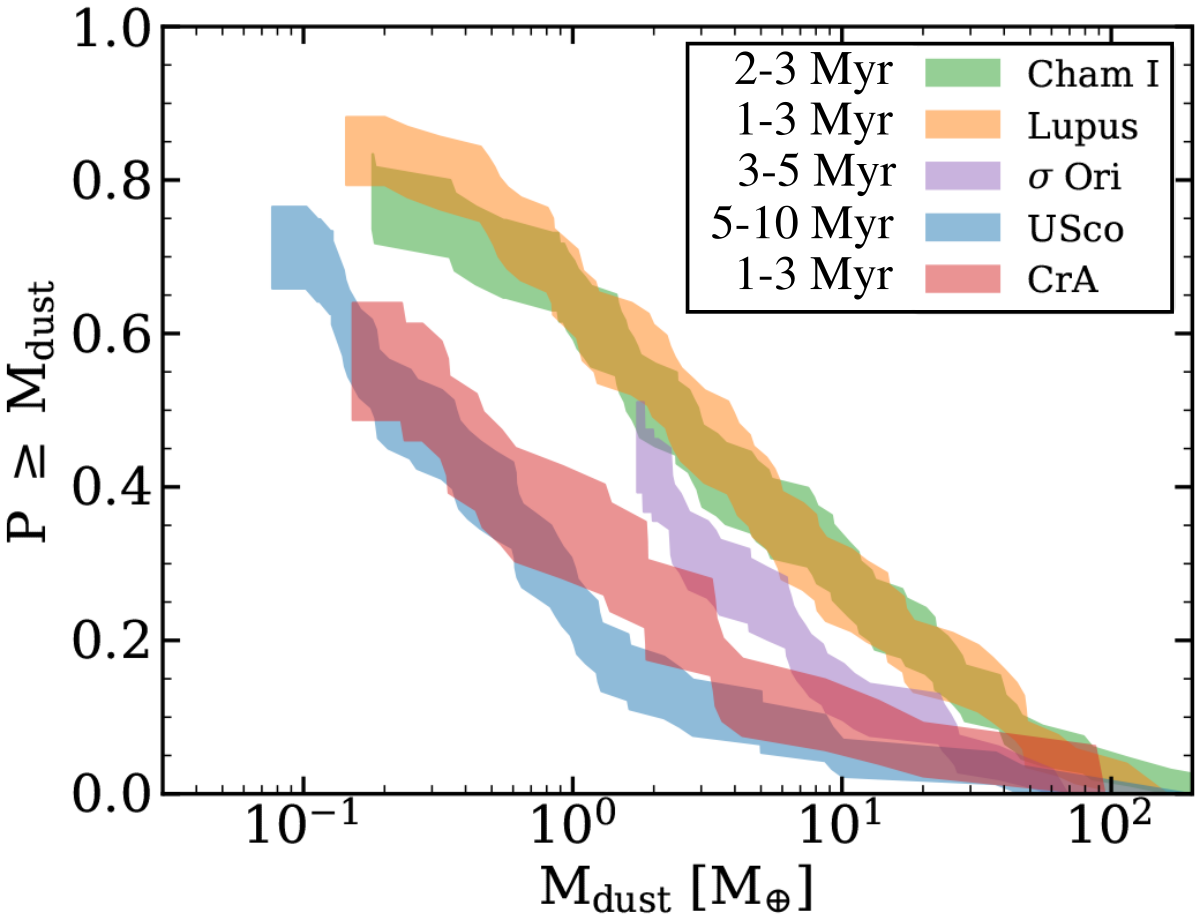}
        \caption{Distribution of disk masses in Chamaelon I (green), Lupus (orange), $\sigma$-Orionis (purple), Upper Sco (blue), and CrA (red). The figure is adapted from Fig. 4 of \cite{Cazzoletti2019}.}
        \label{fig:Cazzoletti}
\end{figure}

However, not all of the observations are consistent with an age-$M_{\rm dust}$ correlation. \citet{Eisner2018} reported significantly lower average disk masses for the $\sim$1 Myr old star-forming region Orion Nebula Cluster (ONC) \citep{Prosser1994,Hillenbrand1997} compared to disks of similar ages in star-forming regions of lower density. 
Consistent with the lower dust mass, \citet{Eisner2018} also measured a significantly smaller average disk radius in the ONC than in Chamaeleon I \citep{Pascucci2016}, Taurus, Ophiuchus \citep{Tripathi2017}, Lupus \citep{Tazzari2017}, and Upper Sco \citep{Barenfeld2017}.  
Similar to the ONC, \citet{Cazzoletti2019} measured an average dust mass of disks in Corona Australis (CrA) of $M_{\rm dust}=6\pm3$ M$_{\odot}$, which is similar to that of Upper Sco (for a comparison of the dust mass distributions of disks in Chamaeleon I, Lupus, $\sigma$-Orionis, Upper Sco, and CrA, see \Fig{Cazzoletti}, which is adopted from \cite{Cazzoletti2019}). 
However, the age of CrA is $\sim1$ -- 3 Myr and hence similar to the ages of Taurus or Ophiuchus, which have much larger and more massive disks. 
Another region with a low average dust mass of disks ($M_{\rm dust}=4\pm1$ M$_{\odot}$), but a young age ($\sim$2 -- 3 Myr) is IC348 \citep{Ruiz-Rodriguez2018}. 
While the presence of lower disk masses in older regions can be explained by radial drift and growth of dust grains, the lower average dust mass of disks in these young regions must have another cause. 

One mechanism leading to smaller disk sizes is disk truncation through tidal interactions with other stars in the star-forming region \citep{ClarkePringle1993,Ostriker1994,Pfalzner2005}.
However, recent simulations have shown that the densities in the considered star-forming regions are too low for frequent and hence efficient disk truncation through tidal stripping \citep{Winter2018a}. 

Another possible mechanism for disk truncation is external photoevaporation by background radiation \citep{StoerzerHollenbach1999,Armitage2000,Adams2010,Facchini2016,Haworth2018}. 
\citet{Winter2018}  compared the effect of tidal stripping with external photoevaporation on protoplanetary disk dispersal. 
They found that the effect of external photoevaporation generally exceeds the effects from tidal interaction, which makes photoevaporation a more likely candidate for mass loss in disks. 
 \citet{Eisner2018} suggested that the massive Trapezium stars associated with the region might be the necessary sources for enhanced external photoevaporation. 
Moreover, $\sigma$-Orionis hosts the massive Herbig star O9V, and the disk sizes correlate with the distance from this Herbig star as shown in observations with Herschel \citep{Mauco2016} and the Atacama Large (sub-)Millimeter Array (ALMA) \citep{Ansdell2017}. 
Therefore, the low disk masses and small disk sizes in $\sigma$-Orionis might be caused by the Herbig star O9V. 
\cite{vanTerwisga2020} detected two distinct disk populations in NGC 2024 with different mean dust masses. 
The eastern population hosts more massive disks that are presumably younger than the disks in the western population. 
The authors interpreted the lower disk masses in the western population as a consequence of both age and enhanced external photoevaporation from IRS 2b and IRS 1 as a result of the lower extinction in the west. 
\cite{vanTerwisga2020} also pointed out that the disks in the western population might have already formed with systematically smaller sizes, as suggested for the disks in the star-forming regions $\rho$ Ophiuchus \citep{Williams2019} and CrA \citep{Cazzoletti2019}.
In particular, CrA hosts a massive star, R CrA, but in contrast to O9V in $\sigma$-Orionis, there is no correlation between disk mass and distance from R CrA. 

Unlike cosmic rays, ultraviolet photons are likely screened soon after leaving the emitting source because of the high extinction within molecular clouds.
Therefore, external photoevaporation is likely not the only mechanism causing smaller disk masses and sizes in young star-forming regions.

A possible explanation for smaller disks and thereby lower disk masses might be the overabundance of low-mass stars in the considered region compared to other regions because the disk mass correlates with stellar mass \citep{Pascucci2016,Ansdell2017}.
This $M_{\rm dust}$ - $M_{\rm \star}$ is a likely explanation for the observation of lower disk masses in IC 348 \citep{Ruiz-Rodriguez2018}.
\citet{Cazzoletti2019} accounted for possible biases in their CrA survey and ruled out that the lower disk masses observed in CrA are caused by the $M_{\rm dust}$ - $M_{\rm \star}$ correlation. 
This means that the smaller disk sizes must have another explanation. 

Surveys of YSO Class 0 and Class I objects show that disks already form early during the star formation process from material falling onto the star-disk system from the surrounding envelope \citep{Jorgensen2009,Segura-Cox2018,Tychoniec2018,Andersen2019}. Disks may therefore already be born with systematic deviations depending on the conditions in the star-forming region.
In this paper, we suggest that at least part of the lower disk masses obtained in recent surveys of Class II objects are a consequence of a higher ionization rate in the star-forming regions with smaller disks. Instead of considering a scenario in which the disks are truncated after their formation, such as it is the case for tidal stripping or external photoevaporation, we suggest that the disks were already born small as a result of stronger magnetic braking. 
It is known from theory that magnetic fields can reduce the sizes of disks significantly or even suppress their formation altogether under idealized conditions \citep[the so-called magnetic braking catastrophe;][]{LuestSchlueter1955,MestelSpitzer1956,MellonLi2008}. 
The  Continuum And Line Young ProtoStellar Object (CALYPSO) survey of Class 0 disks by \citet{Maury2019} suggests that magnetic braking efficiently prevents the formation of large disks during the protostellar collapse phase. 

Today, we know that multiple effects such as turbulence \citep{Seifried2012,Seifried2013,Joos2013,Kuffmeier2017}, misalignment of angular momentum and magnetic field vector \citep{Joos2012,Krumholz2013,Gray2018}, and/or nonideal magnetohydrodynamical (MHD) effects \citep{MellonLi2009,Machida2011,Machida2011b,Tsukamoto2015b,Masson2016,Wurster2019} can circumvent the magnetic braking catastrophe.
In this study, we show in consistence with previous models \citep{Wurster2018CRIRstudy} that for higher ionization rates, magnetic braking is more efficient and leads to lower disk masses.
In other words, nonideal MHD effects, in particular ambipolar diffusion, lead to larger disks for lower ionization rates because magnetic braking depends on the ionization fraction of the gas \citep{Zhao2016}.
Considering that the size of the gaseous disk also determines the size of the dusty disk, we therefore suggest that regions with disks of lower mass, such as the ONC, IC 348, and in particular CrA, are a result of a higher average ionization rate in these regions than in regions with larger disks of similar age. 

We structure the paper as follows. In section 2 we show the effect of the cosmic-ray ionization rate on resistivities and thereby on the disk size. In section 3 we discuss the results in the context of the observed surveys. In section 4 we conclude and discuss  how our scenario may be tested in future observations.

\section{Methods}
The different nonideal MHD effects depend on multiple physical properties such as density, temperature, chemical abundance, and magnetization. \cite{Zhao2016} developed a chemical network that allows for computing resistivities (ambibolar diffusion $\eta_{\rm AD}$, Ohmic dissipation $\eta_{\rm \Omega}$ 
, and Hall effect $\eta_{\rm H}$) for the different physical properties during the collapse phase. 
We computed the resisitivities using their network
to discuss the effect of the cosmic-ray ionization rate on the nonideal MHD resistivities. 

To test the effect of a varying ionization rate on the formation of protostellar disks, we ran 2D simulations with the MHD code \zeustw\ \citep{Krasnopolsky2010} using a spherical grid analogous to the work presented by \citet{Zhao2016}. The simulations account for self-gravity, assume the gas to be isothermal below a critical density of $\rho_{\rm c}=10^{-13} \unit{g}\unit{cm}^{-3}$ , and assume a scaling of thermal pressure with density of $P\propto\rho^{-5/3}$ for densities above $\rho_{\rm c}$.   We carried out 20 simulations assuming a cosmic-ray ionization rate of $\zeta = 10^{-18}\unit{s}^{-1}$, $\zeta = 5\times10^{-18}\unit{s}^{-1}$, $\zeta = 10^{-17}\unit{s}^{-1}$, $\zeta = 5 \times 10^{-17}\unit{s}^{-1}$ , and $\zeta = 10^{-16}\unit{s}^{-1}$ for four combinations of initial rotation and magnetization. 
The end of each simulation was at about 20 kyr after the formation of the central star.  

Our initial condition was a uniform, isolated core with total mass $M_{\rm c}=1.0 \unit{M}_{\odot}$ and a radius of $R_{\rm c}=10^{17}\unit{cm}\approx 6684 \unit{au}$. 
The initial mass density therefore was $\rho_0 = 4.77\times 10^{-19} \unit{g}\unit{cm}^{-3}$ under the assumption of a molecular weight of $\mu = 2.36$.  
We started with a solid-body rotation of the core with an angular speed of $\omega_0=1\times 10^{-13} \unit{s}^{-1}$ or $\omega_0=2\times 10^{-13} \unit{s}^{-1}$, which corresponds to a ratio of rotational to gravitational energy of $\beta_{\rm rot} = 0.025$ or $\beta_{\rm rot} = 0.1$ \citep{Goodman1993}. 
For simplicity, we considered alignment of the magnetic field with the angular momentum vector and started with a uniform magnetic field of $B_0 \approx 2.13\times 10^{-5} \unit{G}$ or $4.25\times 10^{-5} \unit{G}$ , corresponding to a dimensionless mass-to-flux ratio of $\lambda \equiv \frac{M_{\rm c}}{\pi R_{\rm c}^2 B_0} 2\pi \sqrt{G} \approx 5$ or $\lambda \approx 2.5$.
This level of magnetization is consistent with measurements of the Zeeman splitting of OH \citep{TrolandCrutcher2008}.
To compute the resistivities, we used the MRN \citep[Mathis-Rumpl-Nordsieck][]{MRN1977} grain size distribution with a fixed power-law index between the minimum grain size of $0.1 \unit{\mu m}$ and the maximum grain size of $0.25 \unit{\mu m}$ \citep{Zhao2016},
\begin{equation}
     n(a) \propto a^{-3.5}, 
\end{equation}
where $a$ is the grain size and $n$ is the grain number density.

We assumed outflow boundary conditions at the inner boundary of our domain at radius $r_{\rm in}=3\times 10^{13}\unit{cm}\approx2 \unit{AU}$ and at the outer boundary ($r_{\rm out}=10^{17}\unit{cm}$).
Mass flowing into the inner boundary region was collected at the center to represent the growing star.
Our 2D grid ($r$,$\theta$) consisted of $120\times96$ grid points. The grid was uniform in $\theta$ direction and nonuniform in $r$-direction. We logarithmically increased the spacing in $r$-direction outward by a constant factor of $\approx 1.0647$ with a spacing of $\Delta r =0.1 \unit{AU}$ next to the inner boundary.

\section{Ionization, nonideal MHD, magnetic braking, and disk radii}
\begin{figure}
        \centering
    \includegraphics[width=0.49\textwidth]{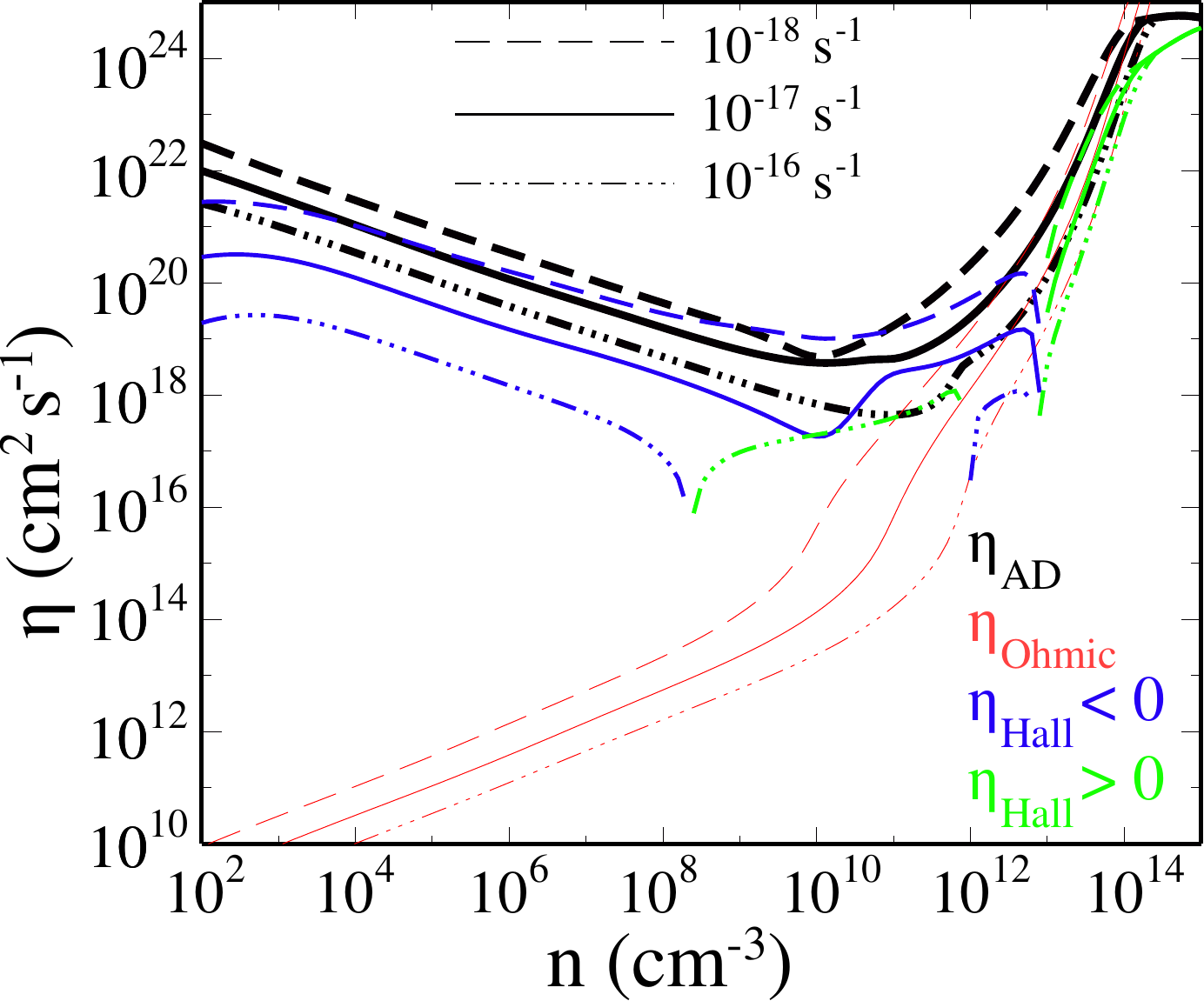}
        \caption{Resistivities (ambipolar: black, Ohmic: red, Hall: green and blue) computed with the chemical network introduced by \citet{Zhao2016} in dependence of the number density of molecular hydrogen H$_2$. 
        Solid lines represent the fiducial case of a cosmic-ray ionization rate of $\zeta = 10^{-17}$ s$^{-1}$, 
        dashed lines represent the case of an ionization rate of $\zeta=10^{-18}$ s$^{-1}$, and dash-dotted lines correspond to $\zeta=10^{-16}$ s$^{-1}$.}
        \label{fig:etadensity}
\end{figure}

\Figure{etadensity} shows the resistivities in dependence of densities computed with the chemical network developed by \citet{Zhao2016}. The resistivities were computed for a magnetic field strength that depends on the hydrogen density as $B(n_{\rm H}) = 1.43\times 10^{-7} \sqrt{ \frac{n_{\rm H}}{\unit{cm}^{-3}}}\unit{G}$ \citep{Nakano2002,Li2011}. The temperature was computed with the broken power law introduced by \citet{Zhao2018} to mimic the results obtained by \citet{Tomida2013} in radiative transfer simulations:

\begin{equation}
  T =
    \begin{cases}
      \footnotesize{\text{for $\rho < 10^{-12}\unit{g}\unit{cm}^{-3}$:} }   \\
      T_0 + 1.5 (\frac{\rho}{10^{-13}\unit{g}\unit{cm}^{-3}}) \\
      \footnotesize{\text{for $10^{-12} \unit{g}\unit{cm}^{-3} \leq \rho \leq 10^{-11} \unit{g}\unit{cm}^{-3}$: } } \\
      (T_0+15 \unit{K}) (\frac{\rho}{10^{-12}\unit{g}\unit{cm}^{-3}})^{0.6} \\
      \footnotesize{\text{for $10^{-11} \unit{g}\unit{cm}^{-3} \leq \rho \leq 3\times 10^{-9} \unit{g}\unit{cm}^{-3}$: } } \\
      10^{0.6} (T_0 + 15 \unit{K}) (\frac{\rho}{10^{-11}\unit{g}\unit{cm}^{-3}})^{0.44} 
    \end{cases}       
,\end{equation}
with mass density $\rho$, temperature $T$ , and $T_0 = 10 \unit{K}$.
We computed the resistivities corresponding to Ohmic dissipation $\eta_{\rm O}$, the Hall effect $\eta_{\rm H}$ , and ambipolar diffusion $\eta_{\rm AD}$ for cosmic-ray ionization rates of $\zeta = 10^{-18}$ s$^{-1}$ (dashed lines), $\zeta = 10^{-17}$ s$^{-1}$ (solid lines), and $\zeta = 10^{-16}$ s$^{-1}$ (dash-dotted lines).
The plot illustrates the high variability of the resistivities and shows that different resistive effects dominate at different densities.

 Ohmic dissipation dominates at densities $n(\mathrm{H}_2) \gtrsim 10^{14}$ cm$^{-3}$ for this particular grain size distribution.
This means that Ohmic dissipation is important at the high densities that are present during the adiabatic phase of the protostar and second core formation \citep[e.g.,][]{Tomida2015,Vaytet2018},
and it also affects the properties in the midplane of the inner disk \citep[e.g.,][]{Gressel2015,Bai2016}.   
In contrast, Ohmic dissipation does not affect the processes beyond $r>10$ au from the star significantly.
For lower densities, $\eta_{\rm H}$ and $\eta_{\rm AD}$ are the dominating resistivities. 

Consistent with theoretical analysis \citep{Hennebelle2016}, comparison runs with and without 
ambipolar diffusion suggest a significant effect of ambipolar diffusion, leading to a plateau value of $B_{\rm rms}\approx 0.1$ G for densities in the range of
$10^{-14}$ g cm$^{-3}$ to about $10^{-8}$ g cm$^{-3}$. The effect is reduced magnetic braking 
and hence enhanced disk formation \citep[see the core collapse simulation by][]{Masson2016}.

The effect of $\eta_{\rm H}$ is the least studied effect of the three resistivities.
In contrast to Ohmic dissipation and ambipolar diffusion, the Hall effect depends on the orientation of the magnetic field
axis with respect to the orientation of the rotational axis. 
In the case of anti-alignment, the Hall effect can potentially increases the rotation of the disk and the resulting disk size, 
whereas alignment of the two axes can lead to stronger magnetic braking and hence smaller disk sizes \citep{Tsukamoto2015b,Marchand2018}.
In this study, we only consider ambipolar and Ohmic resistivities and neglect the Hall effect. 

\Fig{etadensity} shows that the cosmic-ray ionization rates strongly affect the resisitivities in a nonlinear way, with higher ionization rates 
corresponding to significantly lower resitivities. 
Cosmic rays are considered as the main source for ionizing dense clouds, and although variations of the ionization rate caused by focusing and mirroring
at high densities of protostellar collapse have been studied and discussed \citep{PadovaniGalli2011,Silsbee2018}, 
assuming a canonical rate of $10^{-17}$ s$^{-1}$ is common practice in running nonideal MHD simulations \citep[cf.][]{Tomida2015,Vaytet2018}.
This value is often introduced as an overestimate with the argument that the inner parts of the collapse are efficiently shielded from cosmic rays 
and that $^{26}$Al becomes the main source of ionization in the inner part, causing an ionization rate of only $\sim10^{-18}$ s$^{-1}$ \citep{Umebayashi2009}. 
However, observations of dense cores show a wide scatter of cosmic-ray ionization rates $\zeta$ in the range of $\zeta < 10^{-17}$ s$^{-1}$ and 
$\zeta > 10^{-15}$ s$^{-1}$ , with a mean value around $\zeta \sim 10^{-17}$ s$^{-1}$ to $\zeta \sim 10^{-16}$ s$^{-1}$ \citep[see Figure 15 in][with  data from \cite{Caselli1998}]{Padovani2009}.  
Moreover, more recent observations have reported ionization rates of $\zeta\approx3\times 10^{-16}$ s$^{-1}$ in the protostellar shock L1157-B1 \citep{Podio2014}
or even higher rates ($\zeta > 10^{-14}$ s$^{-1}$) around the young protostar OMC-2 FIR 4 \citep{Ceccarelli2014,Fontani2017,Favre2018}. 
These energetic particles produced by the young protostars could increase the ionization rate in the surrounding cloud, affecting the dynamical evolution of contracting dense cores embedded in the irradiated region.

\begin{figure*}
        \centering
        \includegraphics[width=\textwidth]{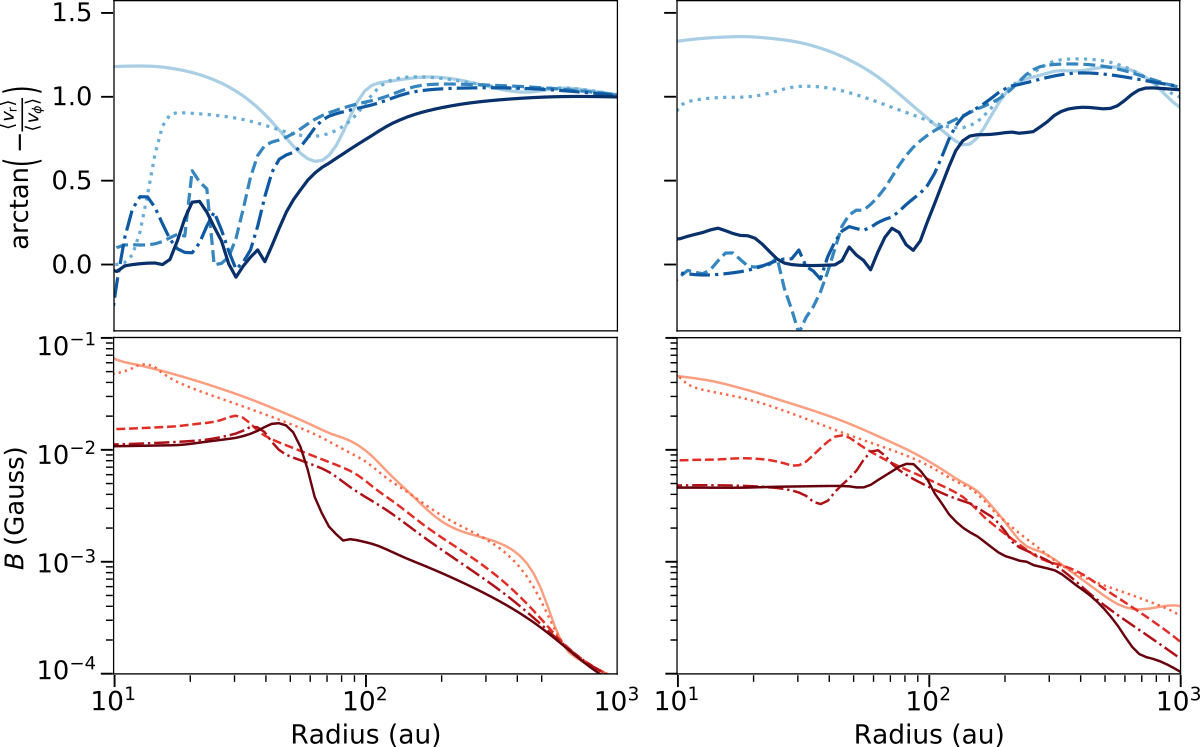}
        \caption{$\alpha=\arctan(-\langle v_{\rm r}\rangle / \langle v_{\phi}\rangle$ (blue lines in the top panels) and magnetic field strength $|\mathbf{B}|$ (red lines in the bottom panels) over radius from the central star for five cosmic-ray ionization rates: $\zeta = 10^{-18} \unit{s}^{-1}$ (dark solid lines), $\zeta = 5 \times 10^{-18} \unit{s}^{-1}$ (dash-dotted lines), $\zeta = 10^{-17} \unit{s}^{-1}$ (dashed lines), $\zeta = 5 \times 10^{-17} \unit{s}^{-1}$ (dotted lines), and $\zeta = 10^{-16} \unit{s}^{-1}$ (light solid lines) . The lighter the color, the higher the ionization rate. The left plots show the radial profiles at $t\approx 5$ kyr, and the right panels at $t\approx 16$ kyr for the combination of $\lambda=4.8$ and $\beta=0.025$.}
        \label{fig:radial-plots}
\end{figure*}

\begin{figure*}
        \centering
        \includegraphics[width=\textwidth]{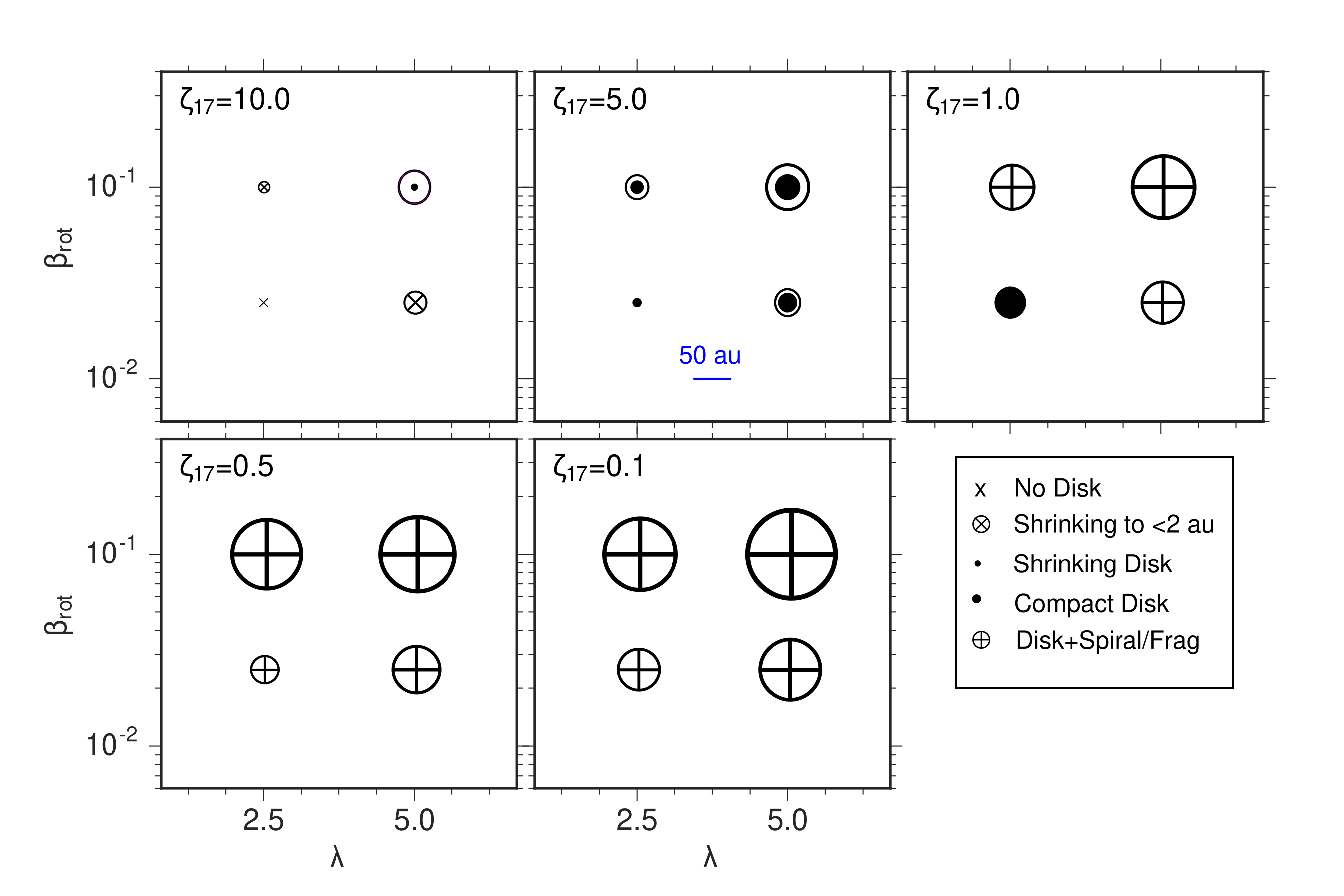}
        \caption{Combined radius of disk and possible spiral structures in dependence of cosmic-ray ionization rates $\zeta$. 
        For reference, the blue bar in the top middle panel indicates the size of a disk with 50 au in diameter. For the shrinking disks, the size of the outer ring denotes size of the initial disk, while the size of the inner bullet denotes the final disk size. For the disk+spiral/frag models, whose spiral structures are growing over time, the symbol size only marks the lower limit of the disk+spiral structures.
        }
        \label{fig:diskradius}
\end{figure*}

\subsection{Disk size dependence on cosmic-ray ionization rate}
The upper panel in \Fig{radial-plots} shows the averaged radial profiles of the radial and azimuthal velocities $v_{\rm r}$ and $v_{\phi}$ displayed as $\alpha = \arctan(-\langle v_{\rm r}\rangle / \langle v_{\phi}\rangle)$ for the combination of $\lambda=4.8$ and $\beta=0.025$ for five ionization rates ($\zeta = 10^{-18} \unit{s}^{-1}$, $\zeta = 5 \times 10^{-18} \unit{s}^{-1}$, $\zeta = 10^{-17} \unit{s}^{-1}$, $\zeta = 5 \times 10^{-17} \unit{s}^{-1}$ , and $\zeta = 10^{-16} \unit{s}^{-1}$) 11 kyr apart in time. $\alpha=\pi/2$ corresponds to pure infall, $\alpha=0$ means pure rotation, and $\alpha=-\pi/2$ is pure outflow.
The darker the line in the plot, the lower the ionization rate. For decreasing ionization rates, the plot shows a trend to lower values of $|\alpha|$, that is, the gas predominantly rotates for low ionization rates, while it predominantly infalls for high ionization rates. 
The trend toward a rotationally dominated profile for decreasing ionization rates correlates with a decrease in magnetic field strength, as illustrated in the lower panels of \Fig{radial-plots}. 
In other words, the increasing resistivity for decreasing ionization rates weakens the transport of angular momentum and hence promotes the formation of rotating structures such as disks. 
As demonstrated in \cite{Zhao2016}, the rotational velocity profile is consistent with Keplerian rotation, corresponding to the accumulated mass of and around the forming protostar at the center.

When we compare the different times with each other, the plot shows a trend: for the highest ionization rates we considered ($\zeta = 10^{-16} \unit{s}^{-1}$ and $\zeta = 5 \times 10^{-17} \unit{s}^{-1}$), the gas becomes even more dominated by infall, while for the lower ionization rates, the radial extent of the rotationally dominated region grows over time. 
Again, the lower panels show that this can be directly attributed to the magnetic field strength, which establishes a characteristic plateau \citep{Hennebelle2016,Masson2016}, but not for the two cases with the highest ionization.    

To summarize the results in a comprehensive form, the five plots in \Fig{diskradius} display the rotational properties for varying ionization rates of $\zeta = 10^{-18} \unit{s}^{-1}$, $\zeta = 5 \times 10^{-18} \unit{s}^{-1}$, $\zeta = 10^{-17} \unit{s}^{-1}$, $\zeta = 5 \times 10^{-17} \unit{s}^{-1}$ , and $\zeta = 10^{-16} \unit{s}^{-1}$ at $t_{\rm end} \approx 20 \unit{kyr}$) of each simulation. 
Each plot shows the properties for four combinations of initial rotation $\beta = 0.1$ or $\beta=0.025$ and mass-to-flux ratio $\lambda=2.5$ or $\lambda=5$ at a fixed ionization rate.
The symbols indicate the characteristics of the rotational profile, including the size and evolution of the disk during the simulation. 
 
Gas belongs to the disk when the density $\rho > 3\times 10^{-13}\unit{g}\unit{cm}^{-3}$, the radial velocity component $v_{\rm r}$ , and the vertical velocity components $v_{\rm z}$ are lower than 50 $\%$ of the azimuthal velocity $v_{\phi}$ ($v_{\rm r} < 0.5 v_{\phi}$ and $v_{\rm z} < 0.5 v_{\phi}$).  
Because we only carried out 2D simulations, what we refer to as a disk may in reality be more spiral in structure in 3D simulations \citep[see][]{Zhao2018}.

\begin{figure*}
        \centering
    \includegraphics[width=\textwidth]{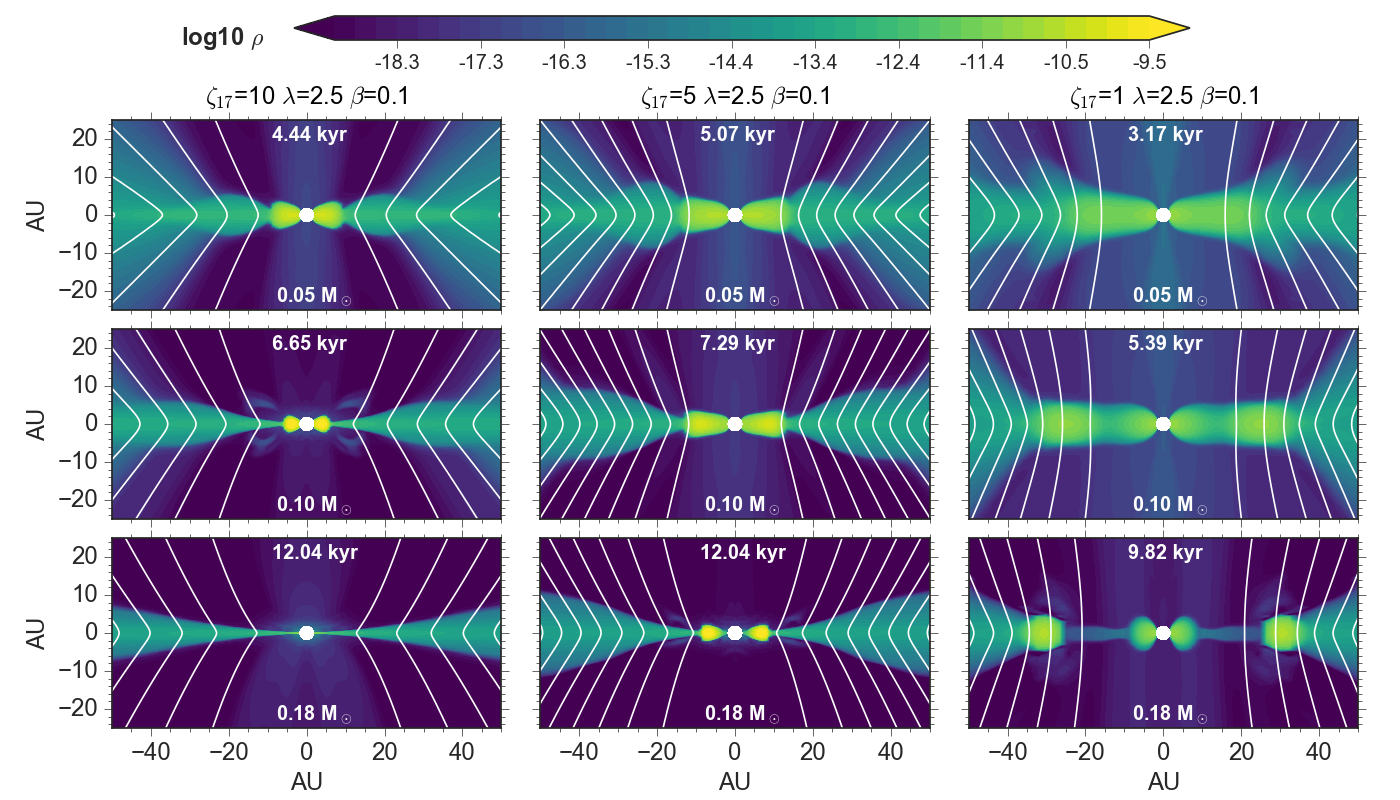}
        \caption{Evolution of the density distribution for $\zeta=10^{-16}$ s$^{-1}$ (left panels), $\zeta=5\times10^{-17}$ s$^{-1}$ (middle panels), and $\zeta=10^{-17}$ s$^{-1}$ (right panels). The initial rotation level is $\beta=0.1$ and the mass-to-flux-ratio is $\lambda=5$ in all of the three cases.     
        }
        \label{fig:rho_disks}
\end{figure*}

To illustrate the formation process in the plane of the 2D model for different levels of ionization, we show in \Fig{rho_disks} the density distribution for identical initial rotation $\beta=0.1$ and magnetization $\lambda=2.5$, but varying ionization rates of $\zeta=10^{-16}\unit{s}^{-1}$ (left panels), $\zeta=5\times10^{-17}\unit{s}^{-1}$ (middle panels), and $\zeta=10^{-17}\unit{s}^{-1}$ (right panels). 
The figure shows the density distribution, when the total mass of the star and the disk has reached a total mass of $0.05$ M$_{\odot}$, $0.1$ M$_{\odot}$ , and $0.18$ M$_{\odot}$. 
For the lowest ionization rate of $\zeta=10^{-17} \unit{s}^{-1}$ shown in \Fig{rho_disks}, the final mass of $0.18$ M$_{\odot}$ is reached earlier at $t\approx 10 \unit{kyr}$ than in the cases of higher ionization ($t\approx 12 \unit{kyr}$) because magnetic pressure support is less efficient. 
In addition, the plots show that for higher ionization levels, the size of the disk shrinks over time. 
In the case of $\zeta=10^{-16} \unit{s}^{-1}$ (left panels), the disk even disappears altogether by the end of the simulation. 
The density distribution corresponding to the canonical ionization rate of $\zeta=10^{-17} \unit{s}^{-1}$ (right panels) show a larger rotating structure more than $30$ au in radius. 
This larger rotational structure would likely correspond to the formation of a spiral structure in a 3D simulation instead of marking the radial extent of a smooth Keplerian disk \cite[see the discussion in][]{Zhao2018} \footnote{We refer to \citet{Zhao2016}, who presented an extensive analysis of other physical properties in the 2D setup for a given cosmic-ray ionization rate ($\zeta=10^{-17}\unit{s}^{-1}$) and discussed the limitations of the model. 
Here, the 2D models serve as a proof of concept, demonstrating the effect of differences in the ionization rate. They are not used as a detailed study of disk properties. }. 
A protostellar companion might form in the spiral in a 3D simulation. 
Likewise, as spiral arms can transport angular momentum outward, the increase in angular momentum at larger radii might lead to an expansion of the disk in radius. 
In any case, even if the actual disks are smaller than the radial extent of the spirals, \Fig{radial-plots} and \Fig{diskradius} show a clear correlation: a decreasing level of ionization corresponds to the formation of larger rotationally supported disks.

\subsection{Variation in initial magnetization and rotation}
In addition to variations in cosmic-ray ionization rate, other parameters are also important for disk formation.
Under the premise of an isolated collapsing core, we investigated the two most prominent effects  for different cosmic-ray ionization rates using the previous
model setup: the amount of angular momentum, and the initial magnetic field strength. 
In the runs displayed in \Fig{rho_disks}, the initial mass-to-flux ratio is $\lambda \approx 2.5$ and the initial rotation of the core is set to $\beta \approx 0.1$. 
For comparison, we carried out simulations for the five ionization rates with the following combinations of $\lambda$ and $\beta$: $\lambda \approx 2.5$ and $\beta \approx 0.025$, $\lambda \approx 5$ and $\beta \approx 0.1$, and $\lambda \approx 5$ and $\beta \approx 0.025$.

\Figure{diskradius} shows that the correlation between decreasing ionization rate and increasing disk size is a robust result for any given combination of rotation $\beta$ and mass-to-flux ratio $\lambda$.
Consistent with previous results by \citet{Zhao2016}, the largest disks form for the combination of high mass-to-flux ratio $\lambda$ and high initial rotation $\beta$ for the collapsing core at a given ionization rate (see \Fig{diskradius}). 
The plot also illustrates that a different ionization rate can lead to disks of similar size if the initial rotation or magnetization of the cloud differs as well. 
For instance, the radial extent of the disk-spiral structure is similar for $\zeta=10^{-17}\unit{s}^{-1}$ with $\beta=0.1$, $\lambda=2.5$ to the case of lower ionization rate $\zeta=10^{-18}\unit{s}^{-1}$, but lower initial rotation $\beta=0.025$.
Nevertheless, the results demonstrate the important effect of the ionization rate on the disk size. 
In the case of the highest ionization rates $\zeta=10^{-16}\unit{s}^{-1}$, even for the most favorable scenario of a maximum initial rotation $\beta=0.1$ and minimum magnetization $\lambda=5$, the disk has a size of 20 au at most, and shrinks as a consequence of continuous magnetic braking to less than 4 au by the end of the simulation.  
In contrast, in the case of the lowest ionization rate $\zeta=10^{-18} \unit{s}^{-1}$ , a disk-spiral structure of 20 to 30
au forms and the disk-spiral structure tends to grow further by the end of
the simulation even for the least favorable combination of low initial rotation $\beta=0.025$ and high magnetization $\lambda=2.5$.  

\subsection{Comparison with other studies}
\citet{Wurster2018CRIRstudy} conducted a similar parameter study of the core-collapse scenario up to densities of about $\rho=10^{-11}$ g cm$^{-3}$ 
. They computed the equations of nonideal MHD as based on different ionization rates 
and compared their results with the ideal MHD case. 
For rates down to $10^{-16}$ s$^{-1}$, the magnetic energy only differs by $\sim 1 \%$ at most and the maximum magnetic field strength differs only up to $10 \%$ 
at densities $\rho<10^{-12}$ g cm$^{-3}$.
When rates of $\zeta > 10^{-14}$ s$^{-1}$ are considered, the properties of maximum density, magnetic energy, and maximum magnetic field strength differ by less
than 1 $\%$ compared to the ideal MHD case. 
We suggest that cosmic-ray ionization rates higher than the canonical value of $\zeta = 10^{-17}$ s$^{-1}$ might be more realistic in many disk-forming regions.
The canonical value originally stems from an early estimate of the minimum cosmic-ray ionization on Earth \citep{SpitzerTomasko1968}.
Based on the suggestion and model presented in \citet{Padovani2016}, 
\citet{GachesOffner2018} recently demonstrated that protostellar accretion shocks may accelerate cosmic rays, which results in cosmic-ray ionization rates of 
$10^{-2}$ s$^{-1}$ to $10^{0}$ s$^{-1}$ at the shock surface and induces higher ionization rates of $\zeta > 10^{-16}$ s$^{-1}$ within the natal cloud of protoclusters.
\citet{Offner2019} even found cosmic-ray ionization rates of $\zeta > 10^{-15}$ s$^{-1}$ on the disk surface. 
However, these high ionization rates of the disk surface occur already after the disk has formed, and therefore the ionization rate during the protostellar collapse is therefore essential for determining the disk size.
The missing bimodality in the distribution of disk radii may indicate that $\eta_{\rm H}$ is lower than $\eta_{\rm AD}$ , consistent with \Fig{etadensity}, and the Hall effect therefore only modestly affects the disk \citep[see results for $\zeta > 10^{-16}$ s$^{-1}$ in][]{Wurster2018inVaytet}.

Considering later stages of disk evolution, \citet{Cleeves2013} suggested that stellar winds reduce cosmic-ray ionization to $\zeta < 10^{-18}$ s$^{-1}$, although \citet{Padovani2018} showed that ionization rates may be significantly higher when a higher flux of stellar protons is assumed. 
Nevertheless, even for very low cosmic-ray ionization rates, we expect a minimum ionization rate in the range of $\sim 10^{-19}$ s$^{-1}$ to $\sim 10^{-18}$ s$^{-1}$ of young protoplanetary disks $<10^6$ yr \citep{Umebayashi2009} even if cosmic-rays were strongly shielded because the short-lived radionuclides $^{26}$Al also ionize the gas in disks of $10^6$ yr.
Measurements of the cosmic-ray ionization rate of $\zeta < 10^{-19} \unit{s}^{-1}$ around TW Hya agree with this hypothesis 
considering the age of TW Hya (up to $\sim 10^7$ yr) and the decay time of $^{26}$Al ($t_{1/2}\approx 7.17\times10^5$ yr). 

However, $\zeta$ can probably be higher in the initial phase of young disks, which are associated with a denser environment of a protocluster. The higher ionization rate would hence lead to the formation of smaller disks than in lower ionized regions, as shown in \Fig{diskradius}.
Considering that the ONC as well as CrA host protostellar clusters, we suggest that these regions are more ionized than Taurus, Lupus, or Ophiucus and that the smaller disks in the ONC or CrA reflect the higher level of ionization in these regions.
Large uncertainties are related to the ionization rates
at the current time. We therefore strongly encourage carrying out observational programs to constrain the ionization rates in denser environments.

\section{Discussion}
\subsection{Limitations of the chemical modeling}
We have discussed the dependence of $\eta$ on density and magnetic field strength. 
Using values computed with the chemical network of \citet{Zhao2016}, 
we also pointed out the dependence of the coefficients on temperature and chemical composition \citep[see also][]{Marchand2016}.
Prior to this study and the work by \citet{Wurster2018CRIRstudy}, the effect of different resistivities has also been considered only for Ohmic dissipation in collapse simulations by \citet{Machida2007} or for Ohmic dissipation and ambipolar diffusion in a simplified setup with a thin-disk approximation by \citet{Dapp2012}. 
However, to ensure that the appropriate resistivities are applied, possibly important nonequilibrium chemistry effects would need to be accounted for and the chemical reactions would need to be computed 'on the fly' during computing time,
for example, by employing the chemical framework \krome\ \citep{Grassi2014}. 
 
Related to this, the resistivities strongly depend on the distribution of grain sizes, and in particular, on the presence of very small grains 
\citep[ VSGs;][]{Dapp2012,Dzyurkevich2017,Zhao2018}.  
With the caveat in mind that the underlying microphysical processes can be very complex \citep{Grassi2019}, it is an appropriate choice to rely on a separate chemical equilibrium model at the current stage, 
also to avoid excessive computing times. 
However, the ionization chemistry converges rather quickly on timescales of $\lesssim 10 \unit{yr}$ \citep{Caselli2002II}, and accounting for nonequilibrium chemistry likely has only little effect.
Generally, a more detailed treatment would not affect the overall result that lower ionization correlates with higher resistivity values. 

\subsection{Limitations of the physical model and setup}
For individual star-forming regions, 
the surveys show that there is not just one characteristic disk radius or disk mass, but a distribution of disks with varying sizes and masses. 
This scatter is likely the result of the variations in prestellar core properties such as turbulence, magnetization, or level of rotation within individual giant molecular clouds (GMCs).
Using zoom-in simulations of star-disk systems forming in a turbulent GMC, \citet{Kuffmeier2017} demonstrated that accretion onto the protostar is a diverse process that leads to the formation of disks of various size even when everywhere in the cloud ideal MHD is considered, that is, full ionization.
Consistently, \citet{Bate2018} also found differences in size and mass of disks in simulations without magnetic fields and hence without ionization. The variations in disk properties that are observed within individual star-forming regions therefore likely reflect the heterogeneity in the accretion and disk formation process \citep[see also][]{Kuffmeier2018,Kuffmeier2019}. 
This would imply that the distribution of disk sizes is more random than is actually observed.
Certainly, the ionization rate alone is not the only parameter that can cause differences in disk size. 
As shown in \Fig{diskradius}, differences in the amount of initial angular momentum or the level of magnetization induce differences in disk size as well. 
Moreover, many other factors might cause differences in disk size, such as turbulence \citep{Seifried2012,Seifried2013}, misalignment of magnetic field and angular momentum vector \citep{Joos2012,Krumholz2013}, different initial density profiles of the collapsing prestellar core \citep{Machida2014}, or differences in the thermal to gravitational energy \citep{Tsukamoto2018}.
These factors are without any doubt also important factors in determining the properties of disks. 
Nevertheless, the result presented in this study of 2D models shows that the level of ionization affects the radius and properties of young disks that are still forming. 

In reality, protostellar collapse and disk formation is a 3D process that occurs in the environment of turbulent GMCs. Additionally, realistic prestellar cores can significantly deviate from the setup of a collapsing isolated sphere. 
For the purpose of illustrating the effect of varying ionization rates on the resulting disk size, the use of 2D models based on the spherical collapse assumption is justifiable.

\subsection{Origin of the observed variations in disks}
The question is whether the differences between the distributions are a result of observational biases that are due to different assumptions or methods. 
Related to this concern, assumptions on  dust temperature, for instance, might be correct for one star-forming region, but different for another.
Nevertheless, although there are uncertainties in measuring disk properties, the general consensus is that the differences between star-forming regions are real. 
It is also reassuring when the same group measures significant differences between star-forming regions (e.g., measurements by \citet{vanTerwisga2019} of disks in the OMC-2, and by \cite{vanTerwisga2020} in the eastern and western part of the NGC 2422.)

Our results show that the net shift toward on average larger or smaller disks can be induced by the variations in ionization between different star-forming regions during the disk formation stage.
Because the ONC as well as CrA host massive protostellar clusters, we expect a higher level of ionization in these regions than in other regions, such as Lupus, Taurus, or Chamaeleon I.

We only modeled the early formation phase, that is, Class 0 in observational terms, while the shift in mean disk mass is predominantly observed for more evolved Class II disks. 
However, the idea that the observed discrepancy between disk sizes in different star-forming region originates from the formation phase rather than from the later evolution is also consistent with surveys of Class 0/I objects in Perseus and Orion. 
Although there is disagreement between surveys regarding the dust masses of disks in Perseus (the VLA/ALMA Nascent Disk and Multiplicity (VANDAM) survey \citep{Tychoniec2018} finds higher masses than the Mass Assembly of Stellar Systems and their Evolution with the SMA (MASSES) survey \citep{Andersen2019}), the results by \citet{Tobin2020} suggest that Class 0/I disks are smaller in Orion than in Perseus. 
Considering that Orion is a more active star-forming region than Perseus, we expect that the result of smaller disks in Orion compared to Perseus is real.  
Finally, we point out that even in the case of a constant but low cosmic-ray ionization rate for different prestellar cores, the level of ionization may still differ by factors of ten because of the differences in the distribution of $^{26}$Al, as shown in hydrodynamical simulations \citep{Vasileiadis2013,Kuffmeier2016,Fujimoto2018}.

We argue that differences in the ionization rates are a promising explanation for a shift of the observed mean disk size between star-forming regions. 
Other factors, such as a deviation in initial angular momentum, magnetization, or dust distribution, might be more relevant for governing the scatter of disk properties within an individual star-forming region we described above. 
Based on our results, we suggest that differences in the mean ionization level of individual star-forming regions are of fundamental importance for the observed systematic difference of the mean disk size between star-forming regions. 

\subsection{Outlook: future models}
The results we obtained from 2D simulations of collapse simulations can certainly only serve as a first proof of concept. The aim for future studies is to constrain the effect of cosmic-ray ionization in more elaborate 3D models with more realistic initial and boundary conditions. 
An essential step toward deeper insight is a more detailed treatment of the propagation of cosmic rays, accounting for attenuation and focusing on the embedded phase of star formation. 
The ultimate goal would be a direct coupling of the resistivities to the ionization rate that is induced by the local cosmic-ray ionization.

\section{Conclusion}
We presented results from a nonideal MHD simulation of a collapsing protostar, accounting for the effects of ambipolar diffusion and Ohmic dissipation. 
In the models, we varied the ionization level to compute the resistivities with a chemical network.
Consistent with the literature, we find that a higher level of ionization corresponds to a higher magnetization around the forming protostar and hence leads to the formation of smaller disks. 
The trend is robust for various combinations of initial rotation and magnetization of the collapsing core. 
Although disks of similar size can form for different ionization rates when the initial rotation or magnetization varies, the runs show that the ionization rate is a powerful regulator of disk formation. Therefore our results suggest that the observed differences in average disk sizes for star-forming regions of similar age are at least in part a consequence of the differences in the ionization level that most likely is induced by cosmic rays. 
In particular, we speculate that the protostellar clusters in the ONC and CrA cause a higher cosmic-ray ionization rate and thereby smaller disks in these regions than in regions such as Taurus, Ophiuchus, Chamaeleon I, or Lupus. 
Differences in disk size already occur during the early phase of star and disk formation, which is consistent with observations of Class 0/I disks in Perseus and Orion \citep{Tobin2020}.
Our study shows the importance of observationally better constraining the cosmic-ray ionization rate in star-forming regions because it is an important parameter for the efficiency of magnetic braking during the disk formation process.

\begin{acknowledgements}
We thank the referee for insightful comments and suggestions that helped to improve the manuscript. 
The research of MK is supported by a research grant of the Independent Research Foundation Denmark (IRFD) (international postdoctoral fellow, project number: 8028-00025B). 
MK acknowledges the support of the DFG Research Unit `Transition Disks' (FOR 2634/1, DU 414/23-1). 
\end{acknowledgements}

\bibliography{general}
\bibliographystyle{aa}

\end{document}